# Enhancing switching in bistable nanomechanical oscillators by geometrical resonance

Ricardo Chacón

*Departamento de Electrónica e Ingeniería Electromecánica, Escuela de Ingenierías Industriales, Universidad de Extremadura, Apartado 382, E-06071 Badajoz, Spain*

Recently[1], Badzey and Mohanty demonstrated the usefulness of suitable amounts of noise in controlling switching in bistable nanomechanical silicon oscillators by stochastic resonance. Their nanomechanical systems consist of beams that are clamped at each end and driven into transverse oscillation with the use of a radiofrequency source which is modulated to control switching of the beams between two stable states. In their conclusions, they emphasized that: "In sharp contrast to most other systems investigated for stochastic resonance, we use a square-wave modulation, as opposed to the canonical sine wave. This is an experimental consideration, because we found that switching was much more easily achieved with a square wave." However, they do not provide any explanation or literature reference for the underlying mechanism[2,3] which greatly reinforces the noise-induced switching. Indeed, the appearance of responses of both symmetric[2] and asymmetric[3] bistable systems subjected to a square-wave modulation with the same shape and period as the modulation (as those found in the two cases considered by Badzey and Mohanty) was explained in terms of geometrical resonance. This wide-ranging notion was introduced as a nonlinear generalization of the usual or frequency resonance[4] and applied theoretically to control temporal[5] and spatiotemporal[6] chaos as well as patterns[7]. To conclude, future nanomechanical memory elements could very well be based on the cooperative effect of the geometrical resonance and stochastic resonance mechanisms to optimally enhance switching. Badzey and Mohanty's system provides for the first time a beautiful demonstration of this profitable cooperation.